\documentclass[conference,10pt]{IEEEtran}
\IEEEoverridecommandlockouts
\usepackage{cite}
\usepackage{amsmath,amssymb,amsfonts, amsthm}
\usepackage{graphicx}
\usepackage{subfigure}
\usepackage{textcomp}
\usepackage{xcolor}
\usepackage{soul,color}
\usepackage{bm}
\usepackage{booktabs}
\usepackage{array,multirow,graphicx}
\usepackage{url}
\usepackage{mdwlist}
\usepackage{neuralnetwork}
\usepackage{siunitx}
\usepackage{balance}

\newtheorem{theorem}{Theorem} 

\DeclareRobustCommand{\IEEEauthorrefmark}[1]{\smash{\textsuperscript{\footnotesize #1}}}

\usepackage[acronym,shortcuts]{glossaries}

\newacronym{5g}{5G}{fifth-generation}

\newacronym{ae}{AE}{autoencoder}

\newacronym{cae}{CAE}{convolutional autoencoder}
\newacronym{cnn}{CNN}{convolutional neural network}

\newacronym{dl}{DL}{deep learning}

\newacronym{glrt}{GLRT}{generalized likelihood ratio test}
\newacronym{gnb}{gNB}{gNodeB}

\newacronym{iq}{IQ}{in-phase quadrature}

\newacronym{mse}{MSE}{mean squared error}

\newacronym{nn}{NN}{neural network}

\newacronym{sdr}{SDR}{software-defined radio}

\newacronym{tdd}{TDD}{time division duplexing}

\newacronym{ue}{UE}{user equipment}

\newacronym{wips}{WIPS}{wireless intrusion prevention systems}

\begin{document}
\title{One-Class Classification as GLRT for\\ Jamming Detection in Private 5G Networks\thanks{This work was partially supported by the German Federal Office for Information Security within the project ADWISOR5G under grant ID 01MO23030B. This work has been partially funded by the European Commission through the Horizon Europe/JU SNS project ROBUST-6G (Grant Agreement no. 101139068).\\ Corresponding author: Matteo Varotto, email: matteo.varotto@h-da.de}}

\author{
\IEEEauthorblockN{Matteo Varotto\IEEEauthorrefmark{1}, Stefan Valentin\IEEEauthorrefmark{1}, Francesco Ardizzon\IEEEauthorrefmark{2}, Samuele Marzotto\IEEEauthorrefmark{2}, and Stefano Tomasin\IEEEauthorrefmark{2}}\smallskip
\IEEEauthorblockA{
\IEEEauthorrefmark{1}Dep. of Computer Science, Darmstadt University of Applied Sciences, Germany\\
\IEEEauthorrefmark{2}Dep. of Information Engineering, University of Padova, Italy 
}
}

\maketitle

\begin{abstract}
\Ac{5g} mobile networks are vulnerable to jamming attacks that may jeopardize valuable applications such as industry automation. In this paper, we propose to analyze radio signals with a dedicated device to detect jamming attacks. We pursue a learning approach, with the detector being a \ac{cnn} implementing a \ac{glrt}. To this end, the \ac{cnn} is trained as a two-class classifier using two datasets: one of real legitimate signals and another generated artificially so that the resulting classifier implements the \ac{glrt}. The artificial dataset is generated mimicking different types of jamming signals. We evaluate the performance of this detector using experimental data obtained from a private \ac{5g} network and several jamming signals, showing the technique's effectiveness in detecting the attacks. 

\end{abstract}


\glsresetall

\section{Introduction}
\Ac{5g} networks have become increasingly important in everyday life scenarios over recent years, because of their technical advances in wireless communications\cite{5g_iot}. Since they also support mission-critical applications such as smart manufacturing or autonomous driving, they should be adequately protected against security attacks.

Nowadays, \ac{wips} monitor the security status of the transmission channel from the link layer up,  aggregating measurements from the different communication layers\cite{jamm_det_1}\cite{jamm_det_2}. Several attackers, however, have learned to hide their malicious behaviors at layer 2 and above. Thus, a recent trend is to exploit the physical layer to provide security services, often relying on machine learning \cite{Hoang2024Physical}. This paper leverages the recent work \cite{paper_matteo} that introduced \ac{dl} to detect jamming attacks. Any device that injects noise into the band used for communication is considered a jammer aiming at making the service unavailable to cellular devices. In this context, jamming and anti-jamming strategies have been recently surveyed in \cite{Jamming2022Pirayesh}.

We consider the \ac{wips} as a one-class classification problem, also called {\em anomaly detection}. Note that the classifier also needs to detect jamming signals that have never been seen before, and on which it may not have been trained. Indeed, assuming a specific attack pattern may even lead to vulnerabilities in the learned detection model that an informed attacker may exploit. However, this constraint makes the design of anti-jamming techniques more challenging. A typical solution of such a one-class classification problem is the \ac{glrt}, which is used in various contexts, e.g., \cite{Ardizzon2024RNN,Crosara2024Worst}. Still, this solution requires the knowledge of the statistics of received signals in legitimate conditions, which may be problematic to obtain, due to the different characteristics of the radio propagation environments where the private networks are deployed. 

In this paper, we frame the \ac{wips} as a one-class classification problem and tackle it by a \ac{glrt} implemented via supervised learning, in particular a \ac{cnn}. As proven in \cite{glrt_st}, under suitable hypotheses, a \ac{dl} model trained with supervised learning can indeed learn the \ac{glrt}, and thus can be used for one-class classification.  Thus, drawing inspiration from \cite{glrt_st}, the detector builds an artificial dataset for the jammer with uniform distribution in the \ac{iq} sample domain and uses it during the training phase of the \ac{dl} model. The accuracy of the trained model is evaluated using samples taken from a real-world jammer, thus modeling the discrepancy between the detector's prior knowledge and the actual attack statistics. 
The trained model performance is compared to the solution of \cite{paper_matteo} that uses \ac{cae}, a \ac{dl} model that implements a full one-class classification problem. The comparison is based on experimental data, where the detector, jammer, and 5G base station are implemented as \acp{sdr}.

After this introduction, Section \ref{sec:secScenario} describes the studied system and assumptions. Section~\ref{sec:dataset} explains the dataset generation. Section~\ref{sec:method} details the models' design for jamming detection. Section~\ref{sec:results} shows the obtained results. Finally, we draw the main conclusions in Section~\ref{sec:conclusions}.

\section{System Model}\label{sec:secScenario}
\subsection{Security Scenario}
\begin{figure}
\includegraphics[width=1\hsize]{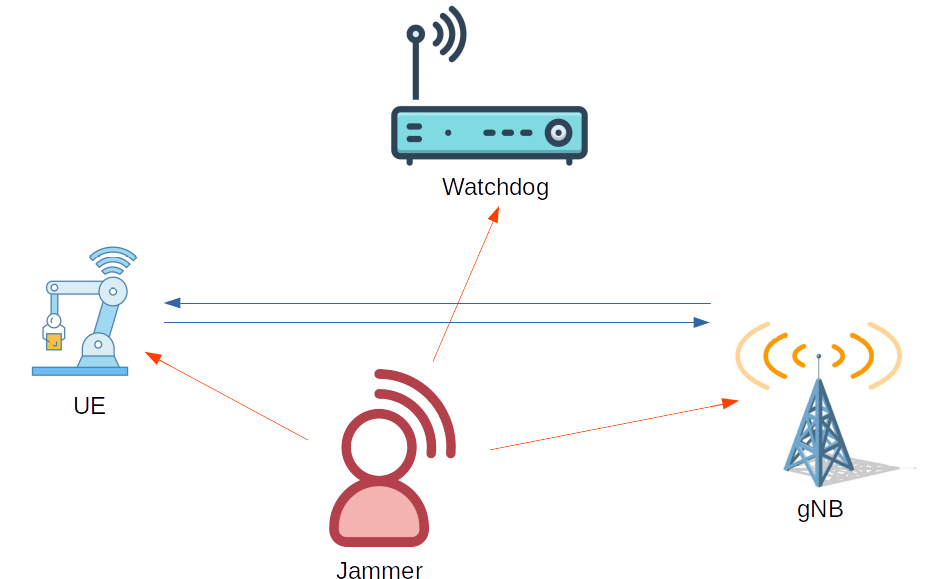}
\caption{Considered security scenario: blue arrows indicate legitimate cellular communications and red arrows indicate the jamming signals.}
\label{scenario}
\end{figure}

We consider the security scenario depicted in Fig.~\ref{scenario} representing a typical private \ac{5g} network used in industrial applications.
We assume that there is at least one wireless channel for communication available, provided by the 5G base station, called \emph{\ac{gnb}}. The wireless channel is used by the \ac{ue} to transmit data whenever necessary.
The availability of the communication is threatened by the \emph{jammer}, which transmits artificial noise in the band allocated for the transmission of data. This band is monitored by the \emph{watchdog}, which permanently records the wireless signal in the form of \ac{iq} samples. Given this stream of \ac{iq} samples, the watchdog aims to detect the attack through a pre-trained machine-learning model.

Adopting the concept of \textit{loose observation}\cite{paper_matteo} we assume that the watchdog knows a priori the basic radio parameters of communication, i.e., carrier frequency  $f_{\rm c}$, bandwidth $W$, and pilot structure.

\subsection{Detector Design via Machine Learning}
In this Section we recall the results of \cite{glrt_st}, detailing how to derive a one-class classifier via machine learning having the same performance as the \ac{glrt}. 

First, we formalize the one-class classification problem. We aim to design a detector to distinguish between a received signal without jamming and one affected by a jamming attack. We consider a scenario where we do not have a database of attack signals, thus the training should be done without prior knowledge of the attack signal.

Formally, let $\mathcal H_0$ be the hypothesis class of no-jamming and $\mathcal H_1$ the hypothesis class of jamming, the jammer detection on a security metric $\Gamma$, which in turn is a function of the input $\bm{X}$ to be tested, is performed as 
\begin{equation}\label{eq:classifier}
\hat{\mathcal H} = \begin{cases}
\mathcal H_0 &\mbox{if } \Gamma(\bm{X}) < \tau, \\
\mathcal H_1 &\mbox{if } \Gamma(\bm{X}) \geq \tau,\\
\end{cases}
\end{equation}
where $\tau$ is the chosen threshold. Thus, we can measure accuracy as the probability of false alarm (FA) and misdetection (MD) defined as
\begin{equation}
P_{\rm FA} = {\mathbb P}[\hat{\mathcal H} = \mathcal H_1 |\mathcal H = \mathcal H_0]\,,
\end{equation}
\begin{equation}
P_{\rm MD} = {\mathbb P}[\hat{\mathcal H} = \mathcal H_0 |\mathcal H = \mathcal H_1]\,.
\end{equation}

A well-known result is that, for a fixed FA, the minimum MD is achieved by using the likelihood ratio
\begin{equation}
    \Gamma_{\rm LR} (\bm{X}) = \frac{p(\bm{X}| \bm{X} \in \mathcal H_0 )}{p(\bm{X}|\bm{X} \in  \mathcal H_1 )}.
\end{equation}
However, such a test requires the knowledge of input statistics when under attack which is not provided in a security scenario, as it would require the cooperation of the attacker during training. Thus, the solution in the statistical framework is resorting to the \ac{glrt},  
\begin{equation}\label{eq:lr}
    \Gamma_{\rm GLRT} (\bm{X}) = {p(\bm{X}| \bm{X} \in \mathcal H_0 )}\,.
\end{equation} 
Such a solution is often used in security applications and is provably optimal in some contexts \cite{Zeitouni92when}. This solution belongs to the statistical domain, where a legitimate dataset distribution is given.

In this work instead, we assume that we only have the dataset $\mathcal{D}_0$ of received signals $\bm{X}$ when operating without jamming. Thus, we have $\mathcal{D}_0 \sim p(\bm{X}| \bm{X} \in \mathcal H_0 )$.

We tackle the one-class classification problem by combining a two-class classifier with an artificial dataset $\mathcal{D}^\star_1$, generated to be uniform over the input domain. The two-class classifier is then implemented as a \ac{cnn}, trained with dataset $\mathcal D = \{\mathcal D_0, \mathcal{D}^\star_1\}$. The following Theorem \cite{glrt_st}[Th. 1] states that such a classifier has the same performance as the \ac{glrt} based classifier, i.e., \eqref{eq:lr} paired with \eqref{eq:classifier}.

\begin{theorem} \cite{glrt_st}[Th. 1]
A \ac{nn} trained with a \ac{mse} loss function over the two class dataset $\mathcal D = \{\mathcal D_0, \mathcal{D}^\star_1\}$, obtain one-class classifiers equivalent to the \ac{glrt}, when a) the training converges to the configuration minimizing the loss functions of the two models, and b) the \ac{nn} is complex enough or the dataset $\mathcal D_0$ is large enough, the training converges to the configuration minimizing the loss functions of the two models.
\end{theorem}

\section{Data Collection and Augmentation}
\label{sec:dataset}
In this section, we will detail how we collected the no-jamming dataset $\mathcal{D}_0$, the jamming dataset $\mathcal{D}_1$ (not used for training, but to assess the performance in testing), and the artificial dataset $\mathcal{D}_1^\star$ (used for training).
We recall that to emulate the one class-classification context, only $\mathcal{D}_0$ and  $\mathcal{D}_1^\star$ have been used for training the detector, while the testing phase is performed between $\mathcal{D}_0$ and $\mathcal{D}_1$.

\subsection{Laboratory Setup}
In the laboratory setup, the devices described above were implemented as \acp{sdr}, specifically:
\begin{itemize}
    \item The \ac{gnb} is implemented based on srsRAN 23.5 \cite{srsran} and the bladeRF 2.0 micro xA4 SDR-frontend\cite{bladerf}. It generates a 5G NR signal with bandwidth $W = \SI{20}{\mega\hertz}$ in the n78 band at center frequency $f_{\rm c}=\SI{3.75}{\giga\hertz}$ in \ac{tdd} mode.
    \item The watchdog and the jammer are implemented based on GNURadio and the ADALM-PLUTO SDR-frontend \cite{plutosdr} and operate at $40$ MHz bandwidth. 
    \item The \ac{ue} is an unmodified \ac{5g} smartphone, namely the Samsung Galaxy A90 (model version: SM-A908B).
\end{itemize}
The core network is provided by Open5GS 2.6.4 \cite{open5gs}.

\subsection{Dataset Creation}
\begin{figure}
     \centering
     \subfigure[\ac{iq} bitmap with no transmission ongoing with time window equal to 1024 samples.]{\includegraphics[width=0.49\columnwidth]{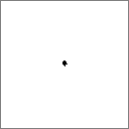}}
     \hfill
     \subfigure[\ac{iq} bitmap with transmission ongoing with time window equal to 1024 samples with an unequalized 4-QAM.]{\includegraphics[width=0.49\columnwidth]{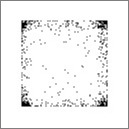}}
\caption{Example of two \ac{iq} bitmaps.}
\label{iq-example}
\end{figure}

Using full bandwidth, the jammer permanently transmits complex noise (uniform and Gaussian), as specified below, while the watchdog permanently records \ac{iq} samples. After having recorded a certain number of \ac{iq} samples, the watchdog creates \ac{iq} bitmaps in a resolution of $128\times 128$ fixing the I and Q axis interval to $[-1.5;1.5]$. Two relevant examples for these bitmaps are given in Fig.~\ref{iq-example} 

\begin{figure}
     \centering
     \subfigure[\ac{iq} bitmap of a uniform distribution of samples over the 2D space with time window equal to 1024 samples.]{\includegraphics[width=0.49\columnwidth]{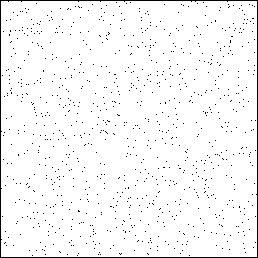}}
     \hfill
     \subfigure[\ac{iq} bitmap of a uniform distribution over a frame with time window equal to 1024 samples.]{\includegraphics[width=0.49\columnwidth]{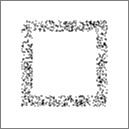}}
\caption{Two types of virtual dataset for training the \ac{dl} models.}
\label{iq-virtual}
\end{figure}

The recorded \ac{iq} data contains the following cases:
\begin{enumerate}
    \item \textbf{Empty channel:} \ac{gnb} actively transmitting beacons but \ac{ue} not transmitting, while the jammer is not active.
    \item \textbf{Transmitting channel}: \ac{ue} and/or \ac{gnb} are transmitting data in TDD mode, while the jammer is not active.
    \item \textbf{Jammer ON}: \ac{ue} and \ac{gnb} occasionally send signals (e.g., beacons, connection requests) but no communication is possible.
    \item \textbf{Artificial attack data:} according to the \ac{glrt} two kinds of synthetic \ac{iq} plots are created: the first one is a uniform distribution of samples over two dimensions (2D), while the second one is a uniform distribution of samples over a frame, as depicted in Fig.~\ref{iq-virtual}. 
\end{enumerate}
IQ bitmaps from the first two cases are classified as legitimate and collected in $\mathcal{D}_0$. Bitmaps from case \emph{Jammer ON} are considered to be in $\mathcal{D}_1$. IQ bitmaps from the last case are placed in the artificial dataset  $\mathcal{D}_1^\star$.

The \emph{number of IQ samples per bitmap} $n$ corresponds to the time window covered per bitmap. This parameter was studied for the levels $n=\{256, 1024, 2048\}$. To save space, we will only present results for $n=256$ in this paper but will comment on the other levels.

\section{Jamming Detection Models}
\label{sec:method}
We adopt two machine learning methods and use them for a systematic comparison as follows: Models built with \ac{cae} will serve as our baseline, since this method only uses data from the not-jammed scenarios, i.e., $\mathcal{D}_0$. \ac{cnn}, instead, will make additional use of the artificial dataset $\mathcal{D}_1^\star$ and, therefore, may or may not outperform the baseline. Let us now describe both methods in detail.

\subsection{Convolutional Neural Network (CNN)}
Our proposed solution includes a CNN, which is a DL model trained with the no-jamming $\mathcal{D}_0$ and the artificial samples $\mathcal{D}_1^\star$. As discussed in Section \ref{sec:secScenario}, such training methods allow the \ac{cnn} jamming detector to achieve the performance of the \ac{glrt} asymptotically.

The designed model is a \ac{cnn}, whose structure is given in Table~\ref{cnn_table}. The CNN is trained to implement \eqref{eq:classifier}, i.e., to return $0$ when $\bm{X} \in \mathcal{D}_0$ and $1$ when $\bm{X} \in \mathcal{D}_1$ (during training, $\mathcal{D}_1^\star$). As a loss function, we adopted the binary cross-entropy
\begin{equation}
L = -\frac{1}{N}\sum_{i=1}^{N}y_{i}\cdot \log(\tilde{y}_{i})+ (1-y_{i})\cdot \log(1-\tilde{y}_{i}),
\end{equation}
where $\tilde{y}_{i}$ is the output for the $i$-th input sample. 
\begin{table}
  \centering
  \renewcommand{\arraystretch}{1}
  \caption{Structure of the Employed \ac{cnn}}  \label{cnn_table}
  \begin{tabular}{c|llr}
    \toprule
    & Layer & Output size & No. of parameters\\
     \midrule
\parbox[t]{2mm}{\multirow{5}{*}{\rotatebox[origin=c]{90}{}}} &     Input              & $128 \times 128 \times 1$ & $0$\\
     & Convolutional 1    & $64 \times 64 \times 64$ & $640$\\
     &Average Pooling   & $32 \times 32 \times 64$ & $0$\\
     &Convolutional 2           & $16\times 16\times 32$ & $18464$\\
     &Flatten        & $8192$ & $0$\\
     &Dense        & $32$ & $262176$\\
     &Dense        & $1$ & $33$\\
     \bottomrule
  \end{tabular}
\end{table}

\subsection{Convolutional Autoencoder (CAE)}
As our baseline, we use the solution of \cite{paper_matteo}. This solution is based on an \ac{cae}, often used for one-class classification \cite{anomaly_det}. An \ac{ae} is a \ac{dl} structure that, given an input $\bm{X}$, compresses it to a latent space with reduced dimensionality (Encoder) and then reconstructs (Decoder) from such compressed representation the original input, outputting $\bm{Y}$. A \ac{cae} is an \ac{ae} that exploits spatial correlation of the 2D data structure by using convolutional filters.

During training, the goal of the model (whose structure is given in Table~\ref{cae_table}) is then to minimize the reconstruction error, defined by the \ac{mse} loss function
\begin{equation}
\bar{\Lambda} = {\mathbb E}[\Gamma], \quad \Lambda(\bm{X}) = ||\bm{X} - \bm{Y}||^2.
\end{equation}

When trained only in the no-jamming dataset, $\mathcal{D}_0$, during testing the \ac{mse} is expected to be small only when $\bm{X} \in \mathcal{D}_0$, while, vice-versa,  the model should output higher \acp{mse} on the jamming cases, $\bm{X} \in \mathcal{D}_1$. The security metric to be used for jamming detection is then $\Gamma_{CAE}(\bm{X}) =\Lambda(\bm{X})$.


\begin{table}
  \centering
  \renewcommand{\arraystretch}{1}
  \caption{Structure of the Employed \ac{cae}}  \label{cae_table}
  \begin{tabular}{c|llr}
    \toprule
    & Layer & Output size & No. of parameters\\
     \midrule
\parbox[t]{2mm}{\multirow{5}{*}{\rotatebox[origin=c]{90}{Encoder}}} &     Input              & $128 \times 128 \times 1$ & $0$\\
     & Convolutional 1    & $64 \times 64 \times 64$ & $640$\\
     &Convolutional 2           & $32\times 32\times 32$ & $18464$\\
     &Flatten              & $32768$ & $0$\\
     &Dense              & $32$ & $1048608$\\
     \midrule
\parbox[t]{2mm}{\multirow{5}{*}{\rotatebox[origin=c]{90}{Decoder}}} &        Input                  & $32$ & $0$\\
&     Dense                  & $32768$ & $1081344$\\
&     Reshape                & $32\times 32\times 32$ & $0$\\
&     Convolutional 1$^T$    & $64\times 64\times 32$ & $9248$\\
&     Convolutional 2$^T$    & $128\times 128\times 64$ & $18496$\\
&     Convolutional   & $128\times 128\times 1$ & $577$\\
    \bottomrule
  \end{tabular}
\end{table}

\section{Numerical Results}
\label{sec:results}
This section will detail the training process for \ac{cae} and \ac{cnn}, describe the used performance metrics, and, finally, discuss the numerical results.

\subsection{Training Process}
The training process of the \ac{cae} was the same as in \cite{paper_matteo}. It was performed using $\mathcal{D}_0$ (without jamming), based on $4000$ \ac{iq} scatter plots. Half of the bitmaps represent empty channels, while the other half represents a busy channel with ongoing transmission. The validation set contains  $600$ \ac{iq} bitmaps, with the same proportional split between empty and busy channels, whose loss was used to automatize an early stopping with patience set to $4$. 

The \ac{cnn} was trained with $4000$ bitmaps from $\mathcal{D}_0$, equally distributed between the two legitimate cases, and $4000$ bitmaps generated with the uniform virtual distribution according to the parameters specified above, collected in the set $\mathcal{D}_1^\star$. The validation set contains $600$ bitmaps with the same distribution as the training set.    
\begin{table}
  \begin{center}
  \caption{Training and Validations Losses for CAE and CNN}  \label{cnn_cae_train_val}
    \begin{tabular}{l|c|r} 
      \textbf{Model} & \textbf{Training} & \textbf{Validation}\\
      \midrule
      CAE & $0.0133$ & $0.0134$\\
      CNN & $9.7\times 10^{-5}$ & $3.8\times 10^{-4}$\\
    \end{tabular}
  \end{center}
\end{table}

The test set was the same for the two models and was composed of $800$ bitmaps: $400$ were taken from legitimate cases, $\mathcal{D}_0$, while the other half was taken from three attacking situations, i.e., a jammer that injects uniform or Gaussian noise over all the spectrum or uniform jamming over a frame, all collected in $\mathcal{D}_1$.


Table \ref{cnn_cae_train_val} shows values for the resulting training and validation loss at the end of the training phase for \ac{cae} and \ac{cnn}. As training and validation loss are close, we can conclude that the trained model did not incur overfitting problems. 

\subsection{Performance Metrics}
We measure performance in terms of false alarm (FA) and misdetection (MD) rates for a variable threshold of the machine learning output between $0$ to $1$. To simplify comparison, the thresholds providing MD and FA rates of $10^{-2}$ are determined for each model. Then, for each pair of MD-FA curves, the distance between the respective MD and FA thresholds will be measured. The resulting distance value per model can then be compared among the models for a quick overview.

\subsection{Performance Results}
\begin{figure}
\includegraphics[width=1\hsize]{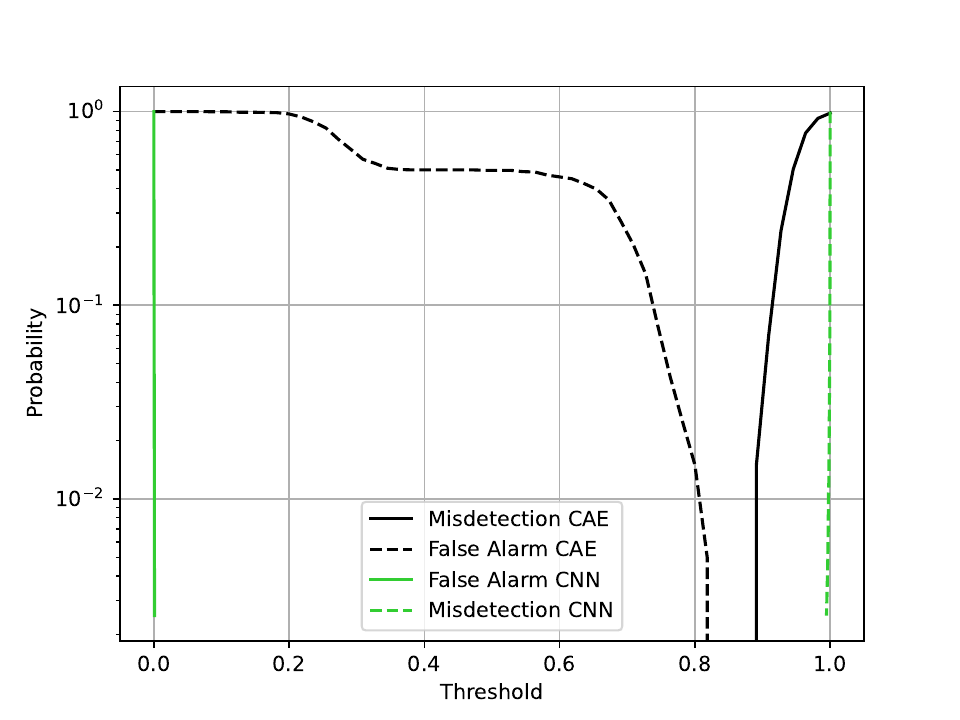}
\caption{Performance comparison in terms of accuracy between the two models in the case: \textit{n}$=256$, noise: uniform.}
\label{comparison_uniform}
\end{figure}

\begin{figure}
\includegraphics[width=1\hsize]{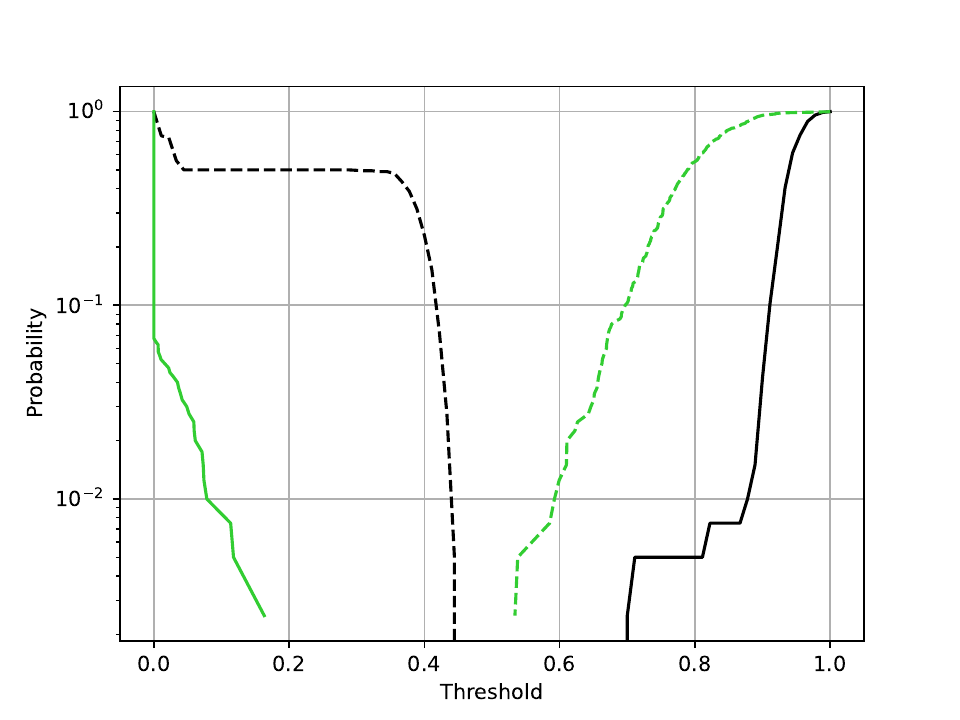}
\caption{Performance comparison in terms of accuracy between the two models in the case \textit{n}$=256$, noise: uniform over a frame. Lines and colors are those of Fig.~\ref{comparison_uniform}.}
\label{comparison_frame}
\end{figure}

\begin{figure}
\includegraphics[width=1\hsize]{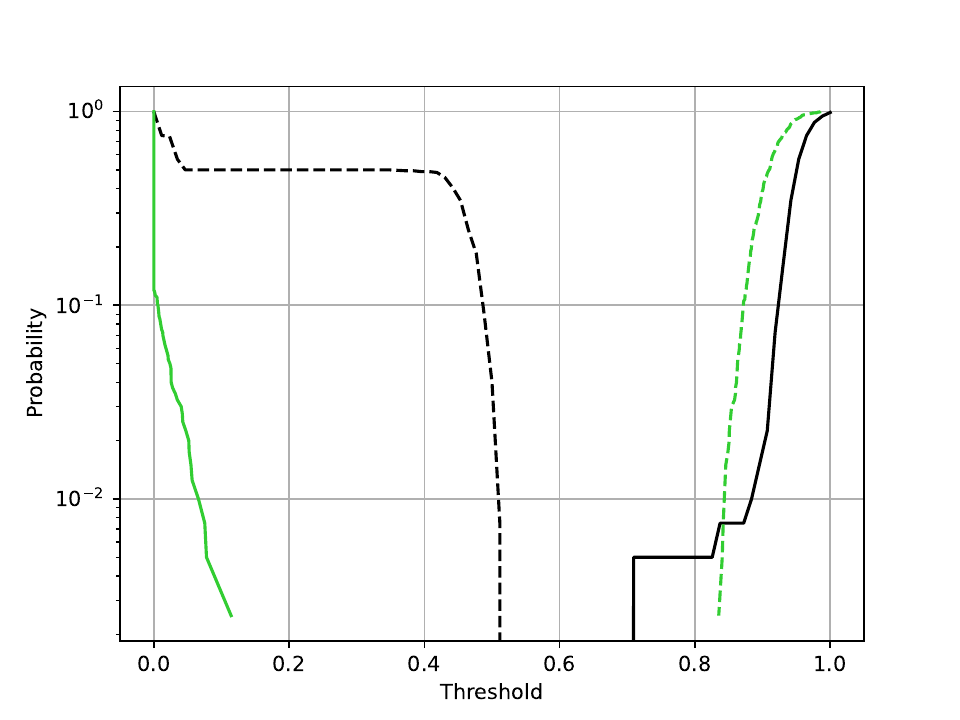}
\caption{Performance comparison in terms of accuracy between the two models in the case \textit{n}$=256$, noise: Gaussian. Lines and colors are those of Fig.~\ref{comparison_uniform}.}
\label{comparison_gaussian}
\end{figure}
Fig.~\ref{comparison_uniform},~\ref{comparison_frame}, and~\ref{comparison_gaussian} compare FA and MD rates achieved with both models for the three different jamming cases.

With uniform noise, the \ac{cnn} clearly shows a better performance than the \ac{cae}.

With frame-like noise (see Fig.~\ref{iq-virtual}b) the \ac{cnn} still outperforms \ac{cae}. This is indicated by the separation between the FA and MD curves for the \ac{cnn}, which is $0.5$ substantially wider than the separation of $0.35$ with CAE. This is a performance gain of $43\%$ over the baseline.

With Gaussian noise, the \ac{cnn} model reaches an even higher performance gain. \ac{cae} produces a separation between the two curves of approximately $0.35$, while the \ac{cnn} produces a separation of approximately $0.75$. Thus, the CNN with artificial data outperforms the baseline by $114\%$.

In addition to $n=256$ IQ samples per bitmap, models created and tested with larger time windows were also studied. With a window size of $n=1024$ samples, the separation of the curves improves, compared to \ac{cae}, but the gain is smaller than with $n=256$. A time window of $2048$ samples, on the other hand, significantly improved separation and gain for the case of uniform noise.
\balance
\section{Conclusions}
\label{sec:conclusions}
For the relevant use case of private 5G networks, we proposed a method to improve the accuracy of a jamming detector at the physical layer.

To keep the detection model independent of the attacker but still profit from supervised learning, we included synthetically generated attack data. The resulting accuracy gains demonstrate improved threat detection compared to the unsupervised learning approach used in previous works.

In light of the promising results and the sensitivity of the subject matter, further studies on the subject will follow.
 
\bibliographystyle{IEEEtran}
\bibliography{biblio.bib}

\end{document}